\documentclass[letter]{aa}    
\usepackage[Symbol]{upgreek}
\usepackage{graphicx}
\usepackage{txfonts}
\usepackage{lipsum}
\usepackage{subcaption}         
\usepackage{lscape}             
\usepackage{placeins}           

\begin{document}

   \title{On-sky dark-hole diggin' with implicit Electric Field Conjugation on MagAO-X\thanks{This paper includes data gathered with the 6.5 meter Magellan Telescopes located at Las Campanas Observatory, Chile.}}


   \titlerunning{On-sky demonstration of iEFC}
   \authorrunning{S. Y. Haffert et al.}

%
%
%

   \author{
   S.~Y.~Haffert\inst{1,2}
   \and
    J.~Liberman\inst{4}
    \and
   J.~R.~Males\inst{2}
    \and
   L.~M.~Close\inst{2}
   \and
   W.~B.~Foster\inst{2}
   \and
   K.~Van Gorkom\inst{2}
   \and
   O. Guyon\inst{2,4,5,6}
    \and
   A.~D.~Hedglen\inst{7}
   \and
   P.~T.~Johnson\inst{2}
   \and
   M.~Y.~Kautz\inst{4}
   \and
   J.~K.~Kueny\inst{4}
   \and
   J.~Li\inst{2}
   \and
    J.~D.~Long\inst{3}
   \and 
   J.~Lumbres \inst{4}
   \and
    M.~Mars \inst{1}
   \and
   E.~A.~McEwen \inst{4}
   \and
   A.~McLeod \inst{8}
   \and
   L. Schatz\inst{9}
   \and
   E.~Tonucci\inst{1}
   \and
   K.~Twitchell\inst{4}
   }

    \institute{Leiden Observatory, Leiden University, PO Box 9513, 2300 RA Leiden, The Netherlands \\
              \email{haffert@strw.leidenuniv.nl}
    \and
    Steward Observatory, The Unversity of Arizona, 933 North Cherry Avenue, Tucson, Arizona
    \and
    Center for Computational Astrophysics, Flatiron Institute, 162 5th Avenue, New York, New York
    \and
    Wyant College of Optical Sciences, The University of Arizona, 1630 E University Blvd, Tucson, Arizona
    \and
    Subaru Telescope, National Observatory of Japan, National Institutes of Natural Sciences, 650 N. A'ohoku Place, Hilo, Hawai'i
    \and
    Astrobiology Center, National Institutes of Natural Sciences, 2-21-1 Osawa, Mitaka, Tokyo, Japan
    \and
    Northrop Grumman Corporation, 600 South Hicks Road, Rolling Meadows, Illinois
    \and
    Draper Laboratory, 555 Technology Square, Cambridge, Massachusetts
    \and
     Starfire Optical Range, Kirtland Air Force Base, Albuquerque, New Mexico
    }

   \date{Received September 30, 20XX}

 
  \abstract
  %
   {Direct spectroscopy is very promising approach to characterizing the atmospheres of nearby rocky exoplanets. Non-common path aberrations (NCPA) are differential aberrations between the science optical path and the adaptive optics optical path. The NCPA leak through the coronagraph and create speckles that mimic exoplanet signals. This limits the sensitivity of high-contrast imaging instruments at close angular separations - exactly the separations where we want to search for rocky exoplanets with current and future telescopes and instruments. }
   {We aim to actively remove the NCPA on-sky during observations by using focal plane wavefront sensing and control with the newly upgraded MagAO-X instrument. }
   {MagAO-X is equipped with a unique second-stage Adaptive Optics (AO) system. The second-stage AO system contains a dedicated deformable mirror (DM) for coronagraphic focal plane wavefront control. This DM is placed after the science and AO beam-splitter and is therefore not seen by the main AO loop. The DM has been recently upgraded from an ALPAO-97 to a Boston Micromachine Kilo-DM. The new Kilo-DM enables focal plane wavefront control with the implicit Electric Field Conjugation (iEFC) algorithm. We developed the necessary procedures to run  iEFC with MagAO-X  on-sky.}
   {We demonstrated the successful removal of NCPA on-sky with an iEFC interaction matrix that was calibrated on the MagAO-X internal source. This demonstrates the repeatability between our off-sky and on-sky alignment. The iEFC algorithm was tested on HR4796A and Alpha Centauri in 0.5” seeing conditions. We saw a reduction of the NCPA by a factor of 2 to 20.}
   {This on-sky validation confirms the robustness and efficiency of iEFC under realistic observing conditions, paving the way for its integration into next-generation AO systems for the Extremely Large Telescope and Giant Magellan Telescope.}


   \maketitle
\nolinenumbers  

\section{Introduction}\label{sec:introduction}
Direct imaging of exoplanets is currently thought to be the most promising pathway to find life in the universe in the near future\citep{borges2024detectability, hardegree2025bioverse}. Over the past decades there have been major advances in high-contrast imaging instruments \citep{kenworthy2025high}. One of the major sources of noise in HCI instruments are quasi-static speckles (QSS). QSS usually are generated inside the instrument itself and slowly evolve over time on timescales that are comparable to the typical observing time. In the end, the QSS do not average out during a typical observing sequence which leads to a strong residual noise level. Active focal plane wavefront control (FPWFC) is a way to remove the QSS and improve the sensitivity limits. Electric Field Conjugation (EFC) in particular is powerful technique for wavefront control. It uses a deformable mirror to inject the exact stellar electric field but with opposite phase to create destructive interference in a chosen region ("the dark hole")\citep{give2007broadband}.

The on-sky demonstration of EFC was achieved only after a prolonged development period, from its inception \citep{give2007broadband,give2011pair} to its eventual demonstration\citep{potier2022increasing}. One of the requirements is a stable and high-Strehl Point Spread Function (PSF). And only now with the second-generation of Adaptive Optics (AO) systems is it possible to use FPWFC techniques to remove the stellar speckles and create the dark region (digging a dark hole) \citep{beuzit2019sphere,males2024magaox}. Furthermore, EFC requires an accurate optical model of the instrument to reconstruct the DM control command. EFC was originally created for space-based HCI instruments such as Roman CGI and HWO. For such systems it is easier to implement EFC because there are fewer optical components compared to ground-based instruments creating a simpler optical model.

A new algorithm called implicit Electric Field Conjugation (iEFC) was created to circumvent the issue of needing an accurate optical model of the instrument \citep{haffert2023implicit}. iEFC uses an empirical calibration of the interaction matrix that can reconstruct the DM command from a measurement of the modulated intensity. The modulated intensity can be measured by multiple methods such as pair-wise probing \citep{give2011pair} or the Self Coherent Camera (SCC) \citep{baudoz2006scc,galicher2010scc}. In practice, iEFC can reach similar contrast levels as EFC in simulations and lab demonstrations \citep{haffert2023implicit, milani2023simulating, desai2024comparative}. Also, due to its flexibility of not needing to know the optical model it can be readily implemented for many other types of optical systems \citep{liberman2024implicit,xin2025implicit}. 

Until now iEFC has only been demonstrated in lab settings. In this letter, we report on the successful first on-sky demonstration of iEFC with the recently upgraded MagAO-X instrument. This is also the first time that a dark hole was created at optical wavelengths on-sky, which is more difficult than at longer Near-Infrared (NIR) wavelengths. The iEFC algorithm was able to reduce the strength of the quasi-static speckles and improve our on-sky contrast. Section 2 describes the setup of MagAO-X and how iEFC was implemented and Section 3 describes the on-sky performance.

\section{The Magellan Adaptive Optics eXtreme (MagAO-X) system} \label{sec:MagAOX}
The Magellan Adaptive Optics eXtreme (MagAO-X) system is an extreme AO instrument specifically developed for high-contrast imaging applications at optical wavelengths. The instrument uses two optical tables that are connected through a periscope system. On the top bench, MagAO-X uses a Woofer-Tweeter architecture with an ALPAO-97 DM (woofer) and a 2K BMC DM (tweeter) to compensate for atmospheric turbulence. This part of the optical system also includes the pupil steering mirror, a tip/tilt mirror in an intermediate focus that can steer the alignment of the pupil on the tweeter DM. The second mirror of the periscope sits on the lower optical bench and is actuated. This can be used to steer the beam and keep it stable on the lower bench during observations. After the periscope system a beam splitter is used to divide the light into the part for the wavefront sensor and the part for the science path.
 
MagAO-X has an unique architecture. A dedicated DM is part of the science path after the wavefront sensor beam splitter. This DM, the Non-Common Path Correction DM (NCPC-DM), is invisible to the AO system allowing it to operate independently from the high-order AO loop. This is very convenient because atmosphere control and (i)EFC have different goals; using both control algorithms with the same DM could lead to competing control commands making it difficult to achieve both goals. MagAO-X solves this by separating the responsibility using a separate DM. Recently, the NCPC-DM has been upgraded from an ALPAO-97 DM to a BMC kilo-DM \citep{kueny2024magao}. The original ALPAO-97 DM was not suitable for EFC because of its low actuator count. The beam is incident on the NCPC-DM under a 45 degree angle. This creates an elliptical beam footprint on the surface of the DM and limits the number of illuminated actuators in vertical direction. An elliptical area of 30 by 24 actuators is illuminated that limits the controllable area in the focal plane to a rectangle of 30$\lambda/D$ by 24 $\lambda/D$. However, controlling both amplitude and phase with a single DM reduces the dark hole area to a one-sided dark hole of 12 by 30 $\lambda/D$.

\subsection{Optical alignment}
\begin{figure}[ht]
\includegraphics[width=\columnwidth]{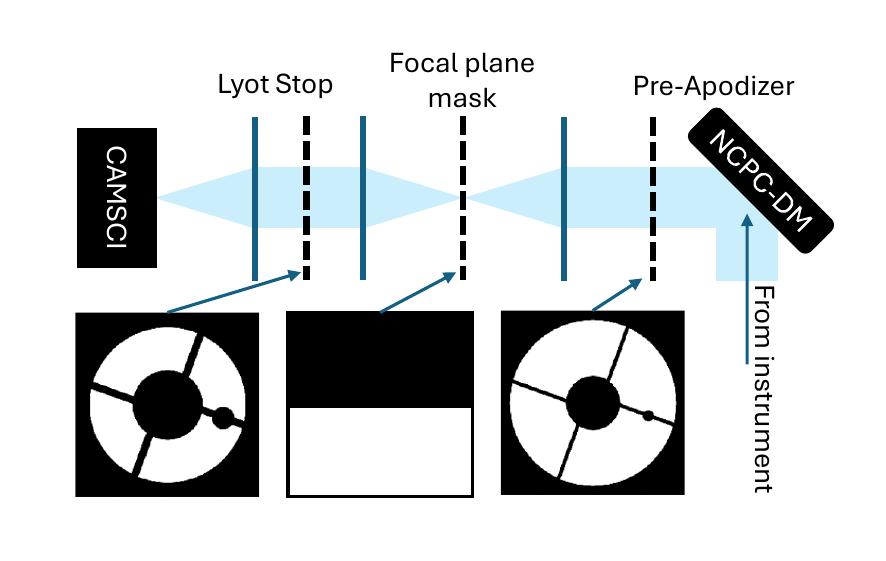}
\caption{A schematic layout of MagAO-X. The beam enters from the bottom right and then first reflects off the NCPC-DM. Afterwards, it goes through the pre-apodizer, the focal plane mask and the Lyot Stop before converging onto the science camera. The white area of the three apodizer masks is transmissive.}
\label{fig:magaox_pupil}
\end{figure}

While iEFC does not require an optical model, it makes the model empirically, it does require the same alignment between the off-sky calibration and the on-sky control. After the NCPC-DM, there are four important optical elements see Figure \ref{fig:magaox_pupil}. The pre-apodizer is used to stop down wavefront errors at the edges from the AO control and mask a bump in the DM surface. The focal plane mask for the work in this letter is a knife-mask which together with the DM apodization from iEFC implements a Phase Apodized Pupil Lyot Coronagraph (PAPLC) \citep{por2020phase}. The final apodizer is the Lyot stop. The beam is then focused onto the science cameras of MagAO-X with an $f/69$ beam. The sampling at 656nm (the H-alpha emission line) is 3 pixels per $\lambda/D$, resulting in a plate scale of 5.97 mas$\mathrm{px}^{-1}$ \citep{long2024astrometric}.

All these elements need to be aligned within certain tolerances to the state that the iEFC interaction matrix was calibrated in. The exact bounds on the tolerances are investigated in a separate work (\cite{liberman2024analyzing}, Liberman et al. in prep.). Several automated alignment procedures are now in use to achieve the required alignment. First, a stage just before the science cameras can insert a lens to create a pupil viewer. This pupil viewer is used to align all the pupil plane elements. The alignment of the pupil w.r.t. the NCPC-DM is found by applying a specifically designed actuator pattern. This actuator pattern defines the zeropoint of our alignment. The pupil is steered to the zeropoint by moving the second periscope mirror. After the pupil alignment, the pre-apodizer is moved into place by the 2-axis stage that allows for x and y control. The offset of the pre-apodizer is found by cross-correlating the current image with the calibration pre-apodizer image. The cross-correlation identifies the apodizer location with $<0.1\%$ precision in terms of pupil diameters. The same alignment strategy is used for the Lyot stop for which a similar alignment precision is achieved.

The alignment strategy for the focal plane mask is less straight forward. The focal plane mask blocks out starlight when aligned. Precise alignment is then difficult especially with atmospheric speckles that leak through the coronagraph mask. A set of incoherent speckles created by the high-order DM are injected at $y=\pm$ 15 $\lambda/D$ and $x=\alpha \lambda/D$ \citep{jovanovic2015artificial}. Here $\alpha$ is an adjustable parameter to set the inner-working angle (IWA) of the coronagraph. The knife-mask does not have a fixed inner-working angle (IWA). It can move up and down to change the IWA. After an IWA is chosen, the focal plane mask is moved until 50\% of the flux of the incoherent speckles is removed.

These procedures align the pupil masks and the focal plane mask and together achieve the required alignment precision. 

\section{Results} \label{sec:results}
In this section we show the on-sky results from iEFC on HR4796A and Alpha Centauri. The atmospheric conditions during the observations were better than median, the seeing at 500 nm was around 0.5" for both objects. The wind speed was between 5 and 10 ms$^{-1}$. These conditions provided a high-quality dataset. Small differences between the beam path of the light from the telescope and our internal telescope simulator requires some small alignment changes on the science cameras. These are on the order of 1 pixel. Even though the PSF shift was only on the order of a pixel, it still has a large effect on the wavefront reconstruction\citep{liberman2024analyzing} making the alignment a necessity to get the algorithm to work. However, the PSF can not be directly tracked because it is blocked by the coronagraph. We use the off-axis incoherent speckles that we create with the high-order DM to track the location of the star. The required PSF shifts are small enough that they can be compensated for by the NCPC DM without impacting the stroke availibility for the iEFC corrections.

\begin{figure}[ht]
\includegraphics[width=\columnwidth]{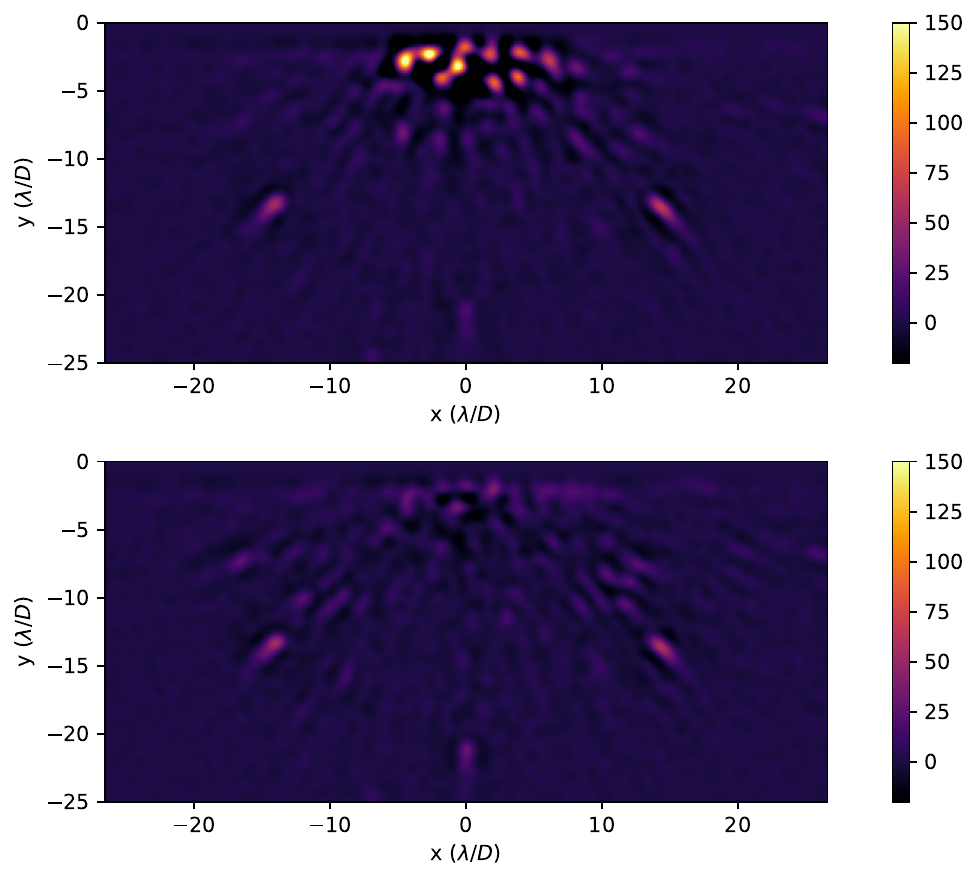}
\caption{High-pass filtered observations of HR4796A before (top) and after (bottom) wavefront control with iEFC. Each image is a combined exposure of 10 seconds. The strong QSS near the inner-working angle are completely removed. The residual speckles that are visible in the images are atmospheric speckle residuals from the wind-driven halo. The speckles at $\pm$ 45 degrees are artificial incoherent speckles created by the high-order DM. The speckle in the south is created by the actuator structure of the NCPC-DM at twice the DM's Nyquist frequency.}
\label{fig:efc}
\end{figure}

We used the z'\footnote{See https://magao-x.org/docs/handbook/observers/filters.html for all definitions of the MagAO-X filters.} filter of MagAO-X for all our observations. The MagAO-X science camera ran at 10 Hz for both Alpha Cen ($I\approx0$) and HR4796 ($I=5.81$) since they are bright objects. A total integration time between 10 and 20 seconds was used per DM probe by stacking the 10 Hz frames. This exposure time range was found to be most stable after several experiments with varying exposure time per probe. Shorter exposure times added more temporal error due to a varying speckle background \citep{singh2019active} and longer exposures were less stable due to outlier events caused by the AO systems. Exposure times between 10 and 20 seconds balanced these two effects and led to a stable closed-loop.

We use single actuator pokes as probes for the iEFC loop. Single actuator probes have been shown to perform very well \citep{potier2020comparing, potier2022increasing,  laginja2025extended}. The single actuator probes were also used for the internal source experiments on MagAO-X \citep{haffert2023implicit}. All on-sky experiments used the same probe amplitude of 0.7 $\upmu$m in surface deviation which is 1.4 $\upmu$m in wavefront deviation. The interaction matrix of the iEFC reconstructor was calibrated on the internal source. The controlled mode basis is a Fourier basis that covered an area of 28 x 10 $\lambda/D$. This is slightly smaller than the theoretical area set by the number of actuators. However, this reduced modal basis was more stable in closed-loop control. Each Fourier mode was calibrated with an amplitude of $\pm$ 15 nm rms in surface deviation. A simple integrator was used for the feedback control with a gain of 0.3. The same gain was used for all modes.

\begin{figure}[ht]
\includegraphics[width=\columnwidth]{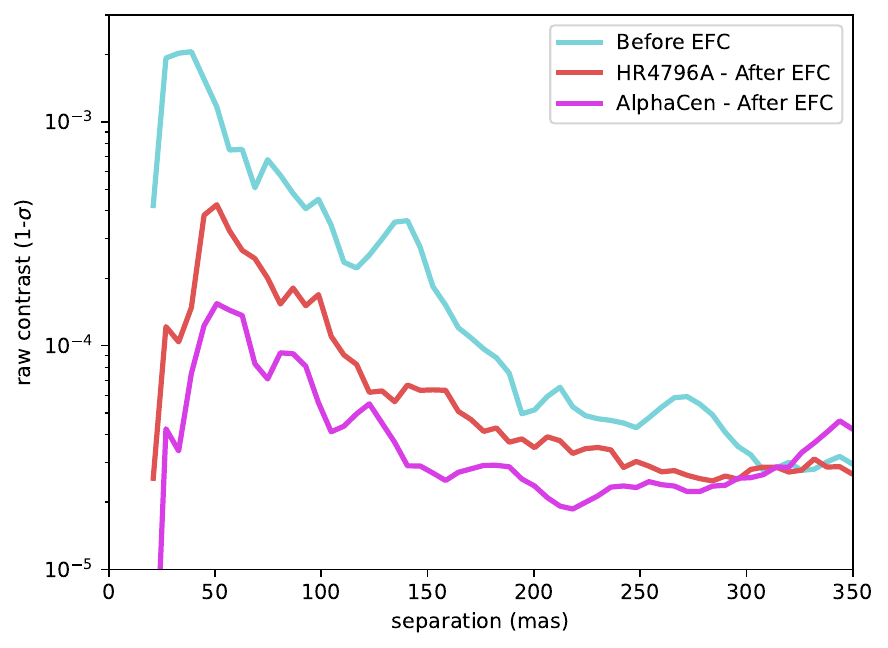}
\caption{The raw contrast curve at z' before and after EFC on two different targets, Alpha Centauri (magenta) and HR4796A (red). The HR4796A curve was made by stacking 60 seconds of exposure time. The Alpha Centauri uses 1 hour of total exposure time. There is an improvement of a factor two to ten depending on the angular separation. The contrast after iEFC is limited by atmospheric residuals.}
\label{fig:efc_contrast}
\end{figure}

Figure \ref{fig:efc} shows the removal of the quasi-static speckles on HR4796A. The total exposure on-target exposure time for HR4796A is roughly 5.0 minutes. The data has been processed using a high-pass filter to filter out global structures such as the wind-driven halo. This enhances the visibility of the QSS. The QSS are significantly reduced especially towards the inner-working angle. The residual speckles are strongly dominated by residual turbulence because the observations were taken at a relatively short wavelength. We also performed the same experiment on Alpha Centauri for which we used a total on target integration time of 138 minutes. The difference in raw contrast is significant. The comparison between the two observations can be seen in Figure \ref{fig:efc_contrast}. The improvement for HR4796A is roughly a factor of 2 to 5 in raw contrast. With more averaging of the atmospheric speckles, we see that on Alpha Centauri the gain is between a factor 2 to 20. The contrast on Alpha Centauri is a factor 2 to 4 deeper than on HR4796A.

\section{Conclusions and Outlook}\label{sec:conclusions}
We presented the first on-sky demonstration of iEFC as a new way of focal plane wavefront control for high-contrast imaging. An important aspect of this demonstration is that we showed that it is possible to do an off-sky calibration and apply the calibration interaction matrix on-sky. This does require a carefully treatment of the alignment. The system needs to be in the same state as during the calibration. The procedures that we developed for MagAO-X are precise enough to achieve this. The algorithm was successfully run on-sky and we reduced the effects of QSS by a factor 2 to 20.

We are limited by an incoherent halo at the inner-working angle that is not purely atmospheric. This is caused by low-order bench seeing. We are currently working on mitigating the bench seeing by reducing instrumental airflow and adding baffles to the optical paths. Preliminary experiments on the internal source indicate that this has improved the post-EFC contrast by almost a factor of 100 at the inner-working angle. We are also working on improving vibration control by using accelerometers to measure the telescope vibrations at extremely high speed (>10 kHz). The measurements will then use a feed-forward control to improve telescope vibration rejection (Johnson et al. in prep). Together with the PIAACMC coronagraph \citep{tonucci2026phase}, these mitigations will allow us to achieve $10^{-5}$ contrast at 1 $\lambda/D$ which we plan to demonstrate in the near future.

The QSS speckles are the dominating noise source for high-contrast imaging at intermediate angular separations. With iEFC on MagAO-X, we can now remove the QSS down to the atmospheric halo limit making that our new limiting noise source. Three approaches are currently being worked on to improve upon this. First, we implemented a  modal gain optimizer \citep{gendron1994astronomical}. This optimizes the feedback gains of the integral controller in-situ during the observations. This will make sure that we will always observe with optimal gain. Secondly, the pyramid wavefront sensor will be run in the unmodulated mode using a non-linear neural network reconstructor \citep{landman2025making}. This will improve the sensitivity of the pyramid wavefront sensor to tip/tilt errors and mid-spatial frequencies. Additionally, due to hardware limitations in the modulator we will also be able to run the AO system at higher frequencies to reduce the servo-lag error that causes the wind-driven halo. Finally, we are investigating predictive control to reduce the servo-lag error \citep{haffert2021data}. All these approaches reduce the strength of the wind-driven halo which is the main source of noise for the iEFC loop. So, not only will better control of the turbulence lead to better raw contrast, the QSS will also be rejected more efficiently.

This initial on-sky validation confirms the robustness and efficiency of iEFC under realistic observing conditions, paving the way for its integration into next-generation AO systems. Future efforts will focus on optimization and assessing long-term stability during extended observing campaigns. The demonstrated performance highlights iEFC’s potential as a key technique for achieving the contrast levels required for direct imaging and characterization of exoplanets with future instruments such as the Planetary Camera and Spectrograph (PCS) for the ELT \citep{kasper2021pcs} and the Giant Magellan Adaptive Optics eXtreme (GMagAO-X) for GMT \citep{males2024high, haffert2024high}.

\begin{acknowledgements}
The MagAO-X phase II project acknowledges generous support from the Heising-Simons Foundation. We are very grateful for support from the NSF MRI Award \#1625441 (MagAO-X). SYH  and MM acknowledge support from NWO Award 184.036.004. JL and SYH acknowledge support from NASA APRA award 80NSSC24K0288. MagAO-X uses the CACAO software package, which is supported by NSF Award \#2410616.
\end{acknowledgements}

\bibliographystyle{aa}
\bibliography{references}

\end{document}